# Scalable photonic reinforcement learning by time-division multiplexing of laser chaos

**Makoto Naruse[1*], Takatomo Mihana[2], Hirokazu Hori[3], Hayato Saigo[4], Kazuya Okamura[5], Mikio Hasegawa[6] and Atsushi Uchida[2]**

[1] *Network System Research Institute, National Institute of Information and Communications Technology, 4-2-1 Nukui-kita, Koganei, Tokyo 184-8795, Japan*

[2] *Department of Information and Computer Sciences, Saitama University, 255 Shimo-Okubo, Sakura-ku, Saitama, Saitama 338-8570, Japan*

[3] *Interdisciplinary Graduate School, University of Yamanashi, Takeda, Kofu, Yamanashi 400-8510, Japan*

[4] *Nagahama Institute of Bio-Science and Technology, 1266 Tamura, Nagahama, Shiga 526-0829, Japan*

[5] *Graduate School of Informatics, Nagoya University, Furo, Chikusa, Nagoya, Aichi 464-8601, Japan*

[6] *Department of Electrical Engineering, Tokyo University of Science, 6-3-1 Niijuku, Katsushika, Tokyo 125-8585, Japan*

* Corresponding author. Email: naruse@nict.go.jp

## Abstract

**Reinforcement learning involves decision making in dynamic and uncertain environments and constitutes a crucial element of artificial intelligence. In our previous work, we experimentally demonstrated that the ultrafast chaotic oscillatory dynamics of lasers can be used to solve the two-armed bandit problem efficiently, which requires decision making concerning a class of difficult trade-offs called the exploration–exploitation dilemma. However, only two selections were employed in that research; thus, the scalability of the laser-chaos-based reinforcement learning should be clarified. In this study, we demonstrated a scalable, pipelined principle of resolving the multi-armed bandit problem by introducing time-division multiplexing of chaotically oscillated ultrafast time-series. The experimental demonstrations in which bandit problems with up to 64 arms were successfully solved are presented in this report. Detailed analyses are also provided that include performance comparisons among laser chaos signals generated in different physical conditions, which coincide with the diffusivity inherent in the time series. This study paves the way for ultrafast reinforcement learning by taking advantage of the ultrahigh bandwidths of light wave and practical enabling technologies.**

Recently, the use of photonics for information processing and artificial intelligence has been intensively studied by exploiting the unique physical attributes of photons. The latest examples include a coherent Ising machine for combinatorial optimization[1], photonic reservoir computing to perform complex time-series predictions[2,3], and ultrafast random number generation using chaotic dynamics in lasers[4,5] in which the ultrahigh bandwidth attributes of light bring novel advantages. Reinforcement learning, also called decision making, is another important branch of research, which involves making decisions promptly and accurately in uncertain, dynamically changing environments[6] and constitutes the foundation of a variety of applications ranging from communication infrastructures[7,8] and robotics[9] to computer gaming[10].

The multi-armed bandit problem (MAB) is known to be a fundamental reinforcement learning problem where the goal is to maximize the total reward from multiple slot machines whose reward probabilities are unknown and could dynamically change[6]. To solve the MAB, it is necessary to explore higher-reward slot machines. However, too much exploration may result in excessive loss, whereas too quick of a decision or insufficient exploration may lead to missing the best machine; thus, there is a trade-off referred to as the exploration–exploitation dilemma[11].

In our previous study, we experimentally demonstrated that the ultrafast chaotic oscillatory dynamics of lasers[2–5] can be used to solve the MAB efficiently[12]. With a chaotic time series generated by a semiconductor laser with a delayed feedback sampled at a maximum rate of 100 GSample/s followed by a digitization mechanism with a variable threshold, ultrafast, adaptive, and accurate decision making was demonstrated. Such ultrafast decision making is unachievable using conventional algorithms on digital computers[11,13,14], which rely on pseudorandom numbers. It was also demonstrated that the decision-making performance is maximized by utilizing an optimal sampling interval that exactly coincides with the negative autocorrelation inherent in the chaotic time series[12]. Moreover, even when assuming that pseudorandom numbers and coloured noise were available in such a high-speed domain, the laser chaos method outperformed these alternatives; that is, chaotic dynamics yields superior decision-making abilities[12].

However, only two options, or slot machines, were employed in the MAB investigated therein; that is, the two-armed bandit problem was studied. A *scalable* principle and technologies toward an *N*-armed bandit with *N* being a natural number are strongly demanded for practical applications. In addition, detailed insights into the relations between the resulting decision-making abilities and properties of chaotic signal trains should be pursued to achieve deeper physical understanding as well as performance optimization at the physical or photonic device level.

In this study, we experimentally demonstrated a *scalable* photonic reinforcement learning principle based on ultrafast chaotic oscillatory dynamics in semiconductor lasers. Taking advantage of the high-bandwidth attributes of chaotic lasers, we incorporated the concept of *time-division multiplexing* into the decision-making strategy; specifically, consecutively sampled chaotic signals are used in the proposed method to determine the identity of the slot machine in a binary digit form.

In the recent literature on photonic decision making, near-field-mediated optical excitation transfer[15,16] and single photon[17,18] methods have been discussed; the former technique involves pursuing the diffraction-limit-free spatial resolution[19], whereas the latter reveals the benefits of the wave–particle duality of single light quanta[20]. A promising approach for achieving scalability by means of near-field-coupled excitation transfer or single photons is *spatial* parallelism; indeed, a hierarchical principle has been successfully demonstrated experimentally in solving the *four*-armed bandit problem using single photons[18]. In contrast, the high-bandwidth attributes of chaotic lasers accommodate *time-division* multiplexing and have been successfully used in optical communications[21]. In this study, we transformed the hierarchical decision-making strategy[18] into the time domain, transcending the barrier toward scalability. We also successfully resolved the bandit problem with up to 64 arms.

Meanwhile, four kinds of chaotic signals experimentally generated in different conditions, as well as quasiperiodic sequences, were subjected to performance comparisons and characterizations, including diffusivity analysis. In addition, computer-generated pseudorandom signals and coloured noise were used to clarify the similarities and differences with respect to chaotically fluctuating random signals. Detailed dependency analysis with regard to the precision of parameter adjustments, sampling interval of chaotic time series, and difficulties of given decision-making problems as well as diffusivity analyses of time series were also performed. The experimental findings will facilitate understanding of the characteristics of laser-chaos-based decision making and the future design of integrated systems.

## Principle

We considered an MAB problem in which a player selects one of $N$ slot machines, where $N = 2^M$ with $M$ being a natural number. The $N$ slot machines are distinguished by the *identity* given by natural numbers ranging from 0 to $N$–1, which are also represented in an $M$-bit binary code given by $S_1 S_2 \cdots S_M$ with $S_i$ ($i$ = 1, …, $M$) being 0 or 1. For example, when $N$ = 8 (or $M$ = 3), the slot machines are numbered by $S_1S_2S_3$ = {000, 001, 010, ..., 111} (Fig. 1a). The *reward probability* of slot machine $i$ is represented by $P_i$ ($i$ = 0, …, $N$–1), and the problem addressed herein is the selection of the machine with the highest reward probability. The reward amount dispensed by each slot machine per play is assumed to be the same in this study. That is, the probability of *winning* by playing slot machine $i$ is $P_i$, and the probability of *losing* by playing slot machine $i$ is $1–P_i$.

The principle consists of the following three steps: **[STEP 1]** decision making for *each bit* of the slot machine in a pipelined manner, **[STEP 2]** playing the selected slot machine, and **[STEP 3]** updating the threshold values. The exact details and general formula are given in the *Methods* section.

### [STEP 1] Decision for *each bit* of the slot machine

The identity of the slot machine to be chosen is determined bit by bit from the most significant bit (MSB) to the least significant bit in a *pipelined* manner. For each of the bits, the decision is made based on a comparison between the measured chaotic signal level and the designated threshold value.

First, the chaotic signal $s(t_1)$ measured at $t = t_1$ is compared to a threshold value denoted as $TH_1$ (Fig. 1b). The output of the comparison is *immediately* the *decision* of the MSB concerning the slot machine to choose. If $s(t_1)$ is less than or equal to the threshold value $TH_1$, the decision is that the MSB of the slot machine to be chosen is 0, which we denote as $D_1 = 0$. Otherwise, the MSB is determined to be 1 ($D_1 = 1$). Here we suppose that $s(t_1) < TH_1$; then, the MSB of the slot machine to be selected is 0.

Based upon the determination of the MSB, the chaotic signal $s(t_2)$ measured at $t = t_2$ is subjected to *another* threshold value denoted by $TH_{2,0}$. The first number in the suffix, 2, means that this threshold is related to the *second*-most significant bit of the slot machine, while the second number of the suffix, 0, indicates that the previous decision, related to the MSB, was 0 ($D_0 = 0$). If $s(t_2)$ is less than or equal to the threshold value $TH_{2,0}$, the decision is that the second-most significant bit of the select slot machine to be chosen is 0 ($D_2 = 0$) (Fig. 1b). Otherwise, the second-most significant bit is determined to be 1 ($D_2 = 1$). Note that the second-most significant bit is determined by the *other* threshold value $TH_{2,1}$ if the MSB is 1 ($D_0 = 1$).

All of the bits are determined in this manner. In general, there are $2^{k-1}$ kinds of threshold values related to the $k$-th bit; hence, there are $2^M - 1 = N - 1$ kinds of threshold values in total. What is important is that the incoming signal sequence is a chaotic time series, which enables efficient exploration of the searching space, as discussed later.

### [STEP 2] Slot machine play

Play the selected slot machine.

### [STEP 3] Threshold values adjustment

Suppose that the selected slot machine yields a reward (i.e. the player *wins* the slot machine play). Then, the threshold values are adjusted so that that same decision will be highly likely to be selected in the subsequent play. Therefore, for example, if the MSB of the selected machine is 0, $TH_1$ should be *increased* because doing so increases the likelihood of obtaining the same decision regarding MSB being 0. All of the other threshold values involved in determining the decision are updated in the same manner.

It should be noted that due to the irregular nature of the incoming chaotic signal, the possibility of choosing the *opposite* values of bits is *not* 0 if the above-described threshold adjustments have been made. This feature is critical in exploration in reinforcement learning. For example, even when the value of $TH_1$ is sufficiently small (indicating that slot machines whose MSBs are 1 are highly likely to be better machines), the probability of the decision to choose machines whose MSBs are 0 is *not* 0. This mechanism is of particular importance when the given decision-making problem is *difficult* (i.e. the differences among the reward probabilities are minute); this situation will be discussed in detail later.

If the selected slot machine does *not* yield a reward (i.e. the player *loses* the slot machine play), then the threshold values are adjusted so that that same decision will *not* be highly likely to be selected in the subsequent play. Therefore, for example, if the MSB of the selected machine is 0, $TH_1$ should be *decreased* because doing so decreases the likelihood of obtaining the same decision regarding MSB being 0. All of the other threshold values involved determining the decision are revised.

As described above, the threshold adjustment involves increasing or decreasing the threshold values based on the betting results, which seems to be *symmetric* between the cases of winning and losing. However, the adjustment must be made *asymmetrically* except in special cases for the following reason.

Suppose that the reward probabilities of Machines 0 and 1 are given by 0.9 and 0.7, respectively, where the probability of receiving a reward is rather high. Indeed, the probability of receiving a reward *regardless* of the decision is (0.9 + 0.7)/2 = 0.8 while that of *no* reward is (0.1 + 0.3)/2 = 0.2. Thus, the event of *losing* is *rare* and should occur *four* times (0.8/0.2 = 4) less than the event of winning. Hence, if the amount of threshold adjustment in the case of winning is set to 1, that in the case of losing should be 4. On the contrary, if the reward probabilities of Machines 0 and 1 are given by 0.1 and 0.3, respectively, the tendency becomes the opposite since most of the betting results in losing; hence, the amount of threshold adjustment in the case of losing must be *attenuated* by four times compared to that in the case of winning.

In the present study, the amount of threshold adjustment in the case of wining is given by 1 while that of losing is given by the parameter Ω. The detailed definition is provided in the *Methods* section. Ω is also updated during the course of play based on the betting history concerning the numbers of *wins* and *selections*. Notably, Ω must be configured differently based on the designated bit. For the MSB, for example, the win/lose events should be related to the two groups of slot machines whose MSBs are 0 and 1, while for the second-most significant bit when the MSB is 0, the win/lose events are related to the two groups of slot machines whose second-most significant bits are 0 and 1 *and* have MSBs of 0.

## Results

A schematic diagram of the laser-chaos-based scalable decision-making system is shown in Fig. 1c. A semiconductor laser operated at a centre wavelength of 1547.785 nm is coupled with a polarization-maintaining (PM) coupler. The light is connected to a variable fibre reflector, which provides delayed optical feedback to the laser, generating laser chaos[22–24]. The output light at the other end of the PM coupler is detected by a high-speed, AC-coupled photodetector through an optical isolator (ISO) and optical attenuator. The signal is sampled by a high-speed digital oscilloscope at a rate of 100 GSample/s (a 10 ps sampling interval) with an eight-bit resolution; the signal level takes integer values ranging from −127 to 128. The details of the experimental setup are described in the *Methods* section.

Figure 2a shows examples of the chaotic signal trains. Four kinds of chaotic signal trains were generated, which are referred to as (i) Chaos 1, (ii) Chaos 2, (iii) Chaos 3, and (iv) Chaos 4 in Fig. 2a, by varying the reflection by the variable reflector by letting 210, 120, 80, and 45 μW of optical power be fed back to the laser, respectively. A quasiperiodic signal train was also generated, as depicted in Fig. 2a(v), by the variable reflector by providing a feedback optical power of 15 μW. Figure 2b summarizes the experimentally observed radio-frequency (RF) power spectra obtained using Chaos 1, 2, 3, and 4 and quasiperiodic signals. It can be seen that the chaotic time series contain wide bands of signals[24] and that there are clear differences among the shapes of the RF spectra corresponding to Chaos 1–4, even though the time-domain waveforms shown in Fig. 2a(i–iv) look similar. The experimental details of the RF spectrum evaluation are provided in the *Methods* section.

In addition, Fig. 2a(vii) shows an example of a coloured noise signal train containing negative autocorrelation calculated using a computer based on the Ornstein–Uhlenbeck process using white Gaussian noise and a low-pass filter[25] with a cut-off frequency of 10 GHz[12]. Also, the black curve in Fig. 2a(vi) marked with RAND depicts a sequence generated by a pseudorandom generator based on the Mersenne Twister. For RAND, the horizontal axis of Fig. 2a should be read as 'cycles' instead of physical time, but we dealt with RAND as if it were available at the same sampling rate as the laser chaos signals to investigate the performance differences between the laser chaos sequences and pseudorandom numbers both in qualitative and quantitative ways.

## Two-armed bandit

We began with the two-armed bandit problem, which is the simplest case[12]. The slot machine was played 250 times consecutively, and such play was repeated 10,000 times. The reward probabilities of the two slot machines, referred to as Machines 0 and 1, were 0.9 and 0.7, respectively; hence the *correct* decision was to choose Machine 0 because it was the machine with the higher reward probability. The red, green, blue, and cyan curves in Fig. 2c show the evolution of the *correct decision ratio* (CDR), defined as the ratio of the number of times when the selected machine has the highest reward probability at cycle $t$ based on the time series of Chaos 1, 2, 3, and 4, respectively. The chaotic signal was sampled every 50 ps; that is, a single cycle corresponds to 50 ps in physical time. The magenta, yellow, and black curves in Fig. 2c represent the CDRs obtained based on quasiperiodic, coloured noise, and RAND sequences. Clearly, the chaotic sequences approach a CDR of unity more quickly than the other signals. Although the difference is subtle, Chaos 3 exhibits the best adaptation among the four chaotic time series; a CDR of 0.95 was achieved at cycle 122, corresponding to 6.1 ns.

In the previous study, an exact coincidence between the autocorrelation of the laser chaos signal trains and the resulting decision-making performance was obtained[12]; specifically, it was found that the sampling interval yielding the negative maximum of the autocorrelation provided the fastest decision-making abilities. To solve a two-armed bandit problem, a single threshold ($TH_1$) and single chaotic signal sample are needed to derive a decision ($D_1$ = 0 or 1). The sampling interval, or more precisely the *inter-decision sampling interval*, of chaotic signals to configure the threshold ($TH_1$) is defined by $\Delta_S$, which is shown in Fig. 1b. Figure 2d compares the autocorrelations of Chaos 1–4 as well as the quasiperiodic and coloured noise. Chaos 1–4 exhibit negative maxima at time lags of around 5 and 6 (and −5 and −6), whereas the quasiperiodic and coloured noise yield negative maxima at time lags of around 7 (and −7). The amount of time lag corresponds to the physical time difference multiplied times 10 ps, which is the sampling interval; hence, for example, a time lag of 5 means that the time difference is 50 ps.

Correspondingly, Fig. 3 characterizes the CDRs as a function of the inter-decision sampling interval $\Delta_S$ by setting the reward probabilities of the two slot machines to 0.1 and 0.5. In Fig. 3a, the CDRs at cycle 10 are compared among Chaos 1–4, while Fig. 3b shows the CDRs at cycle 100 for the quasiperiodic, coloured noise, RAND, and Chaos 3 series. In Fig. 3a, the CDRs obtained using the chaotic time series show maxima around the sampling intervals of 50 ps and 60 ps, which nicely coincide with the negative maxima of the autocorrelations, as we reported previously[12]. At the same time, the negative maxima of the chaotic time series follow the order Chaos 4, 3, 2, and 1 in Fig. 2d, whereas the greater decision-making performances follow the order Chaos 3, 2, 4, and 1 in Fig. 3a with a sampling interval of 50 ps. That is, the order of the absolute values of the autocorrelation does *not* explain the resulting decision-making performances. We will discuss the relation between the decision-making performance and the characteristics of chaotic time series via *other* metrics at the end of the paper. Meanwhile, the black curve in Fig. 3b, which corresponds to RAND, does not show dependency on the inter-decision sampling interval, whereas the magenta and yellow curves corresponding to quasiperiodic and coloured noise exhibit peaky characteristics with respect to the sampling interval, clearly indicating the qualitative differences between correlated times series and conventional pseudorandom signals.

### Multi-armed bandit

We applied the proposed time-division multiplexing decision-making strategy to bandit problems with more than four arms. Here, we first describe the problem to be solved and the assignment of reward probabilities (Fig. 4a).

(1) **Two-armed:** The reward probabilities of Machines 0 and 1 are given by 0.9 and 0.7, respectively (Fig. 4a(i)). Note that the difference is 0.2, which is retained in the subsequent settings.

(2) **Four-armed**: In addition to the threshold used to determine the MSB ($TH_1$), two more thresholds are necessary to determine the second bit ($TH_{2,D_1}$ ($D_1$ = {0, 1})). The reward probabilities of Machines 0, 1, 2, and 3 are defined as 0.7, 0.5, 0.9, and 0.1, respectively, where the correct decision is to select Machine 3 (Fig. 4a(ii)). Note that the difference between the highest and second-highest reward probabilities is 0.2, as in the two-armed bandit problem. In addition, the sum of the reward probabilities of the first two machines (Machines 0 and 1: 0.7 + 0.5 = 1.2) is *larger* than that of the second two machines (Machines 2 and 3: 0.9 + 0.1 = 1.0). This situation is called *contradictory*[18] since the maximum-reward-probability machine (Machine 3) belongs to the latter group whose reward-probability sum is greater than that of the former group.

**(3) Eight-armed:** In addition to the thresholds used to determine the MSB ($TH_1$) and the second bit $TH_{2,D_1}$ ($D_1$ = {0, 1}), four more thresholds are needed to decide the third bit ($TH_{3,D_1,D_2}$ ($D_1$ = {0, 1}, $D_2$ = {0, 1})). The reward probabilities of Machines 0, 1, 2, 3, 4, 5, 6, and 7 are given by 0.7, 0.5, 0.9, 0.1, 0.7, 0.5, 0.7, and 0.5, respectively. First, the difference between the highest and second-highest reward probabilities is 0.2, as in the two- and four-armed bandit problems described above. Second, the sum of the reward probabilities of the slot machines whose MSBs are 0 and 1 are 2.2 and 2.4, respectively, whereas the maximum-reward-probability machine (Machine 2) has an MSB of 0, which is a contradictory situation. Similarly, the sums of the reward probabilities of the slot machines whose second MSBs are 0 and 1 (as well as whose MSBs are 1) are 1.2 and 1, respectively, while the best machine belongs to the latter group, which is also a contradiction (Fig. 4a(iii)). In the following bandit problem definitions, all of these contradictory conditions are satisfied for the sake of coherent comparison with the increased arm numbers.

(4) **16-armed:** In addition to the thresholds used to determine the MSB ($TH_1$), the second bit $TH_{2,D_1}$ ($D_1$ = {0, 1}), and the third bit ( $TH_{3,D_1,D_2}$ ( $TH_{3,D_1,D_2}$ ($D_1$ = {0, 1}, $D_2$ = {0, 1})), eight more thresholds are required for the fourth bit ( $TH_{4,D_1,D_2,D_3}$ ( $TH_{3,D_1,D_2}$ ($D_1$ = {0, 1}, $D_2$ = {0, 1}, $D_3$ = {0, 1})). The reward probabilities of Machines 0, 1, 2, 3, 4, 5, 6, 7, 8, 9, 10, 11, 12, 13, 14, and 15 are given by 0.7, 0.5, 0.9, 0.1, 0.7, 0.5, 0.7, 0.5, 0.7, 0.5, 0.7, 0.5, 0.7, 0.5, 0.7, and 0.5, respectively. The best machine is Machine 2. The contradiction conditions are satisfied, as in the four- and eight-armed problems (Fig. 4a(iv)).

(5) **32-armed:** A 32-armed bandit requires thresholds to determine five bits. The best machine is Machine 2. The contradiction rules apply, as in the previous cases. The details are described in the *Methods* section (Fig. 4a(v)).

(6) **64-armed:** A 64-armed bandit requires thresholds to determine six bits. The best machine is Machine 2. The contradiction rules apply, as in the previous cases. The details are described in the *Methods* section (Fig. aA(vi)).

Figures 4c(i), (ii), (iii), (iv), (v), and (vi) summarize the results of the two-, four-, eight-, 16-, 32-, and 64-armed bandit problems, respectively. The red, green, blue, and cyan curves show the CDR evolution obtained using Chaos 1, 2, 3, and 4, respectively, while the magenta, black, and yellow curves depict the evolution obtained using quasiperiodic, RAND, and coloured noise, respectively. The threshold values take integer values ranging from −128 to 128. The sampling interval of the chaotic signal trains for the MSB ($\Delta_S$) is 50 ps, whereas that of the subsequent bits, called the *inter-bit sampling interval* ($\Delta_L$), is 100 ps. (The impacts due to the choice of $\Delta_L$ are discussed later.) From Fig. 4c, it can be seen that Chaos 3 provides the promptest adaptation to the unity value of the CDR, whereas the nonchaotic signals (quasiperiodic, RAND, and coloured noise) yield substantially deteriorated performances, especially in bandit problems with more than 16 arms. The number of cycles necessary to obtain the correct decision increases as the number of bandits increases. The square marks in Fig. 4d indicate the numbers of cycles required to reach a CDR of 0.95 as a function of the number of slot machines, where the required number of cycles grows in the form of the power-law relation $aN^b$, where *a* and *b* are approximately 55 and 1.16, respectively. These results support the successful operation of the proposed scalable decision-making principle using laser-generated chaotic time series.

## Discussion

### Inter-bit sampling interval dependencies

In resolving MAB problems in which the number of bandits is greater than four and is given by $2^M$, $M$ samples are needed with the interval being specified by $\Delta_L$, as schematically shown in Fig. 1b. In this study, we investigated the $\Delta_L$ dependency by analysing the four kinds of four-armed bandit problems shown in Fig. 4b and labelled as Types 1–4. The reward probabilities of Type 1 are equal to those in the case shown in Fig. 4a(ii); $P_0$, $P_1$, $P_2$, and $P_3$ are given by 0.7, 0.5, 0.9, and 0.1, respectively. The correct decision is to select Machine 2; that is, the machine identity is given by $S_1S_2 = 10$. In deriving the correct decision, the first sample should be *greater* than the threshold ($TH_0$) to obtain the decision $S_1 = 1$, whereas the second sample should be *smaller* than the threshold ($TH_{2,1}$) to obtain the decision $S_2 = 0$. Consequently, if $\Delta_L$ is 0, the search for the best selections does *not* work well since the *same* sampling provides the same searching traces that do not satisfy the conditions for both bits. Indeed, the cyan circular marks in Fig. 5a characterize the CDR at cycle 100 as a function of $\Delta_L$, where $\Delta_L = 0$ ps yields a CDR of 0. Chaos 3 was used for the evaluation. The CDR exhibits the maximum value when $\Delta_L = 50$ ps, which is reasonable because 50 ps is the interval that provides the negative autocorrelation that easily allows oppositely arranged bits to be found ($S_1S_2 = 10$).

Types 2, 3, and 4 contain the same reward probability values but in different arrangements. In Type 3, the correct decision is to select Machine 1, or $S_1S_2 = 01$, which is similar to the correct decision in Type 1 in the sense that the two bits have *opposite* values. Consequently, the inter-bit-interval dependence, shown by the yellow circular marks in Fig. 5a, exhibits traces similar to those of Type 1, where $\Delta_L = 0$ ps gives a CDR of 0, whereas $\Delta_L$ values that yield negative autocorrelations provide greater CDRs. In Types 2 and 4, on the other hand, the correct decisions are given by Machines 0 and 3, or $S_1S_2 = 00$ and $S_1S_2 = 11$. For such problems, $\Delta_L = 0$ ps gives a *greater* CDR due to the eventually *identical* values of the first and second bits, whereas $\Delta_L$ values corresponding to negative autocorrelations yield *poorer* performance, unlike for Types 1 and 3, as clearly represented by the magenta and black circular marks in Fig. 5a.

It is also noteworthy that pseudorandom numbers provide no characteristic responses with respect to the inter-bit intervals, as shown by the square marks in Fig. 5a, which is another clear indication that the temporal structure inherent in chaotic signal trains affects the decision-making performance.

The decision-making system must deal with all of these types of problems, namely, all kinds of bit combinations; thus, temporal structures, such as positive and negative autocorrelations, may lead to inappropriate consequences. To derive a *moderate* setting, the circular marks in Fig. 5b show the *coefficient of variation* (CV), which is defined as the ratio of the standard deviation to the mean value, for the four types of problems shown in Fig. 5a. A smaller CV is preferred. The inter-bit sampling interval of 30 ps eventually provides the *minimum* CV, although slight changes could lead to larger CVs. Indeed, the autocorrelation is about 0 with this inter-bit sampling interval (Fig. 2d). An inter-bit sampling interval of approximately 100 ps constantly offers smaller CV values. The square marks in Fig. 5b correspond to RAND, and *no* evident inter-bit interval dependency related to the CV is observable, in clear contrast to the chaotic time series cases.

## Decision-making difficulties

The adjustment precision of the thresholds is important when searching for the maximum-reward-probability machine, especially in many-armed bandit problems that include the contradictory arrangement discussed earlier[18]. Here, we discuss the dependency of the *decision-making difficulty* by focussing on the two-armed bandit problem; even in simple two-armed cases, the threshold precision clearly affects the resulting decision-making performance.

Figures 6a and 6b present two-armed bandit problems in which the reward probability of Machine 0 is 0.9 whereas that of Machine 1 is 0.5 and 0.7, respectively. Since the probability difference is larger in the former case, it is *easier* to derive the maximum-reward-probability machine in that case. Indeed, the curves in the former case shown in Fig. 6a provide steeper adaptation than in the latter case depicted in Fig. 6b. The eight curves shown therein depict the CDRs corresponding to numbers of threshold levels given by $2^K+1$, where $K$ takes integer values from 1 to 8. From Figs. 6a and 6b, it can be seen that the CDR is *saturated* before approaching unity when the number of threshold levels is smaller than it is in the $K = 5$ case; in particular, the CDR is limited to significantly lower values in difficult problems (Fig. 6a).

One of the reasons behind such phenomena is insufficient exploration due to the smaller threshold levels. In the case of nine-level thresholding ($K = 3$), the estimated reward probabilities of Machines 0 and 1 based on Eq. (6) at cycle 200 are 0.742 and 0.307, respectively, which significantly differ from the *true* values (0.9 and 0.7, respectively). Eventually, $\Omega_1$ became approximately 1.1, which is far from the value of $\Omega_1 = 4$ based on Eq. (7) assuming true reward probabilities, indicating that the threshold could not be biased toward the positive or negative maximum value (here it should be a *positive* maximum because the correct decision is to choose Machine 0); indeed, the threshold $TH_1$ is limited to around −45 at cycle 200, which is far from the negative maximum of −128. Based on Eqs. (1) and (5), the absolute value of the threshold is decreased due to the forgetting parameter, which is 0.45; this value is larger than the average of the decrement and increment caused by the first terms of Eqs. (1) and (5), leading to the saturation of $TH_1$ and resulting in a limited CSR. Figure 6c summarizes the CDR at cycle 200 as a function of decision difficulty, where precise threshold control is necessary to obtain a higher CDR, especially in difficult problems.

## Diffusivity and decision-making performance

In the results shown for bandit problems with up to 64 arms in Fig. 4, Chaos 3 provides the best performance among the four kinds of chaotic time series. The negative autocorrelation indeed affects the decision-making ability, as discussed in Fig. 5; however, the value of the negative maximum of the autocorrelation shown in Fig. 2b does not coincide with the order of performance superiority, indicating the necessity of further insights into the underlying mechanisms.

In this respect, we analysed the diffusivity of the temporal sequences based on the ensemble averages of the time-averaged mean square displacements (ETMSDs)[26,27] in the following manner. We first generated a random walker via comparison between the chaotic time series. If the value of the random number, which was generated based on the Mersenne Twister, is smaller than $s(t)$, the walker moves to the right. $X(t) = +1$; otherwise, $X(t) = -1$. Hence, the position of the walker at time $t$ is given by $x(t) = X(1) + X(2) + \ldots + X(t)$. We then calculate the ETMSD using

$$ETMSD(\tau) = \left\langle \frac{1}{T-\tau} \sum_{t=1}^{T-\tau} \left( x(t+\tau) - x(t) \right)^2 \right\rangle, \quad (1)$$

where $x(t)$ is the time series, $T$ is the last sample to be evaluated, and $\langle \cdots \rangle$ denotes the ensemble average over different sequences. The ETMSDs corresponding to Chaos 1, 2, 3, and 4 and quasiperiodic, RAND, and coloured noise are shown in the inset of Fig. 7, all of which monotonically increase as a function of the time difference $\tau$. It should be noted that at $\tau = 1000$, Chaos 3 exhibits the maximum ETMSD, followed by Chaos 2, 1, and 4, as shown by the circular marks in Fig. 7a. This order agrees with the superiority order of the decision-making performance in the 64-armed bandit problem shown in Fig. 4c. At the same time, RAND derives an ETMSD of 1000 at $\tau = 1000$, which is a natural consequence considering the fact that the mean square displacement of a random walk is given by $\left\langle \left( x(t) - \langle x(t) \rangle \right)^2 \right\rangle = 4pqt$, where $p$ and $q$ are the probabilities of flight to the right and left, respectively. If $p = q = 1/2$ (via RAND), then the mean square displacement is $t$. From Fig. 7a, RAND and coloured noise actually exhibit larger ETMSD values than Chaos 1–4, although the decision-making abilities are considerably poorer for RAND and coloured noise, implying that the ETMSD alone cannot perfectly explain the performances.

Figure 7b explains diffusivity in another way, where the average displacement $\langle x(t) \rangle$ and $\langle x(t+D) \rangle$ are plotted for each time series superimposed in the $XY$ plane with $D$ = 10,000. Although the quasiperiodic and coloured noise, shown by the magenta and yellow curves, respectively, move toward positions far from the Cartesian origin, their trajectories are biased toward limited coverage in the plane. Meanwhile, the trajectories of the chaotic time series cover wider areas, as shown by the red, green, blue, and cyan curves. The trajectories generated via RAND, shown by the black curve, remain near the origin.

To quantify such differences, we evaluated the covariance matrix $\Theta = \mathrm{cov}(X_1, X_2)$ by substituting $x(t)$ and $x(t+\mathrm{D})$ for $X_1$ and $X_2$, where the $ij$-element of $\Theta$ is defined by $\frac{1}{N-1}\sum_{i=1}^{N}\left(X_1-\overline{X_1}\right)\left(X_2-\overline{X_2}\right)$, with $N$ denoting the number of samples and $\overline{X_i}$ denoting the average of $X_i$. The *condition number* of $\Theta$, which is the ratio of the maximum singular value to the minimum singular value[28,29], indicates the uniformity of the sample distribution. A larger condition number means that the trajectories are skewed toward a particular orientation, whereas a condition number closer to unity indicates uniformly distributed data. The square marks in Fig. 7a show the calculated condition numbers, where Chaos 1–4 achieve smaller values, which are even smaller than that achieved by RAND, and the quasiperiodic and coloured noise yield larger scores.

Through these analyses using the ETMDSs and condition numbers related to the diffusivity of the time series, a clear correlation between the greater diffusion properties inherent in laser-generated chaotic time series and the superiority in the resulting decision-making ability is observable.

Simultaneously, however, we consider further insight to be necessary to draw *general* conclusions regarding the origin of the superiority of chaotic time series in the proposed decision-making principle. For example, as can be seen in Fig. 2b, the RF spectrum differs among the chaotic time series, which suggests a potential relation to the resulting decision-making performances. Ultimately, an artificially constructed optimal chaotic time series that provides the best decision-making ability could be derived, which is an important and interesting topic requiring future study.

## Pipelined processing

The design, implementation, and performance analysis were performed offline in this study since the primary objectives were to verify the scalability of the principle and seek better chaotic sequences by physically tuning the original laser chaos. Although we demonstrated scalability up to 64 arms, no greater numbers of arms were employed due to the technological limitations regarding the computing power required to emulate all of the slot machines and the external environment. Simultaneously, however, it should be noted that the proposed decision-making principle and architecture have simple structures with a particular emphasis on pipelined processing[30]. **(1)** The decision of the first bit ($S_1$) of the slot machine depends only on the first threshold ($TH_1$) and first sampled data. No other information is required. **(2)** The decision of the second bit ($S_2$) depends on the decision of the first bit ($D_1$) obtained in the previous step, the second threshold ($TH_{2,D_1}$), and the second sampled data. Simultaneously, the decision of the first bit can proceed to the next decision by sampling the next signal. **(3)** The same architecture continues until the $M$-th bit. The earlier stages do not depend on the results obtained in the latter stages. Such a structure is particularly preferred due to the benefits of the ultrahigh-speed chaotic time series signals and greater throughput of the total system.

## Conclusion

We proposed a scalable principle of ultrafast reinforcement learning or decision making using chaotic time series generated by lasers. We experimentally demonstrated that multi-armed bandit problems with $N = 2^M$ arms can be successfully solved using $M$ points of signal sampling from the laser chaos and comparison to multiple thresholds. Bandit problems with up to 64 arms were successfully solved even though the presence of difficulties that we call contradictions can potentially lead to trapping in local minima. Based on the experimental results, the required latency scales as $N^{1.16}$, with $N$ being the number of slot machines or bandits. Furthermore, by physically changing the laser chaos operation conditions, four kinds of chaotic time series were subjected to the decision-making analysis; a particular chaos sequence provided superiority over the other chaotic time series. Diffusivity analyses through the ETMSDs and covariance matrix condition numbers related to the time sequences well accounted for the underlying mechanisms for quasiperiodic sequences and computer-generated pseudorandom numbers and coloured noise. This study is the first demonstration of photonic reinforcement learning with scalability to larger decision problems and paves the way for new applications of chaotic lasers in the realm of artificial intelligence.

## Methods

### Optical system

The laser was a distributed feedback semiconductor laser mounted on a butterfly package with optical fibre pigtails (NTT Electronics, KELD1C5GAAA). The injection current of the semiconductor laser was set to 58.5 mA (5.37$I_{th}$), where the lasing threshold $I_{th}$ was 10.9 mA. The relaxation oscillation frequency of the laser was 6.5 GHz, and its temperature was maintained at 294.83 K. The optical output power was 13.2 mW. The laser was connected to a variable fibre reflector through a fibre coupler, where a fraction of light was reflected back to the laser, generating high-frequency chaotic oscillations of optical intensity[22–24]. The length of the fibre between the laser and reflector was 4.55 m, corresponding to a feedback delay time (round trip) of 43.8 ns. PM fibres were used for all of the optical fibre components. The optical signal was detected by a photodetector (New Focus, 1474-A, 38 GHz bandwidth) and sampled using a digital oscilloscope (Tektronics, DPO73304D, 33 GHz bandwidth, 100 GSample/s, eight-bit vertical resolution). The RF spectrum of the laser was measured by an RF spectrum analyser (Agilent, N9010A-544) up to 44 GHz every 44 MHz. The observed raw data were subjected to moving averaging over a 20-point window, yielding the RF spectrum curves shown in Fig. 2b.

## Details of the principle

### 1. Decision of the *most* significant bit

The chaotic signal $s(t_1)$ measured at $t = t_1$ is compared to a *threshold value* denoted as $TH_1$ (Fig. 1b). The output of the comparison is *immediately* the *decision* of the *most significant bit* (MSB) concerning the slot machine to choose. If $s(t_1)$ is less than or equal to the threshold value $TH_1$, the decision is that the MSB of the select slot machine to be chosen is 0, which we denote as $D_1 = 0$. Otherwise, the MSB is determined to be 1 ($D_1 = 1$).

### 2. Decision of the *second*-most significant bit

Suppose that $s(t_1) < TH_1$; then, the MSB of the slot machine to be selected is 0. The chaotic signal $s(t_2)$ measured at $t = t_2$ is subjected to *another* threshold value denoted by $TH_{2,0}$. The first number in the suffix, 2, means that this threshold is related to the *second*-most significant bit of the slot machine, while the second number of the suffix, 0, indicates that the previous decision, related to the MSB, was 0 ($D_0 = 0$). If $s(t_2)$ is less than or equal to the threshold value $TH_{2,0}$, the decision is that the second-most significant bit of the select slot machine to be chosen is 0 ($D_2 = 0$). Otherwise, the second-most significant bit is determined to be 1 ($D_2 = 1$).

### 3. Decision of the *least* significant bit

Suppose that $s(t_2) > TH_{2,0}$; then, the second-most significant bit of the slot machine to be selected is 1. In such a case, the *third* comparison with regard to the chaotic signal $s(t_3)$ measured at $t = t_3$ is performed using *another* threshold adjuster value denoted by $TH_{3,0,1}$. The 3 in the subscript 3,0,1 indicates that the threshold is related to the third-most significant bit, and the second and third numbers, 0 and 1, indicate that the most and second-most significant bits were determined to be 0 and 1, respectively. Such threshold comparisons continue until all $M$ bits of information $(D_1, D_2, \cdots, D_M)$ that specify the slot machine have been determined. If $M = 3$, the result of the third comparison corresponds to the least significant bit of the slot machine to be chosen. Suppose that the result of the comparison is $s(t_3) < TH_{3,0,1}$; then, the third bit is 0 ($D_3 = 0$). Finally, the decision is to select the slot machine with $D_1D_2D_3 = 010$; that is, the slot machine to be chosen is 2.

In general, there are $2^{k-1}$ kinds of threshold values related to the $k$-th bit; hence, there are $2^M - 1 = N - 1$ kinds of threshold values in total.

## 4. Threshold values adjustment

### I. When *winning* the slot machine play

If the identity of the selected slot machine is $D_1D_2 \cdots D_M$ and it yields a reward (i.e. the player *wins* the slot machine play), then the threshold values are updated in the following manner.

### (1) MSB

The threshold value related to the MSB of the slot machine is updated according to

$$TH_1(t+1) = +\Delta + \alpha TH_1(t) \quad \text{if } D_1 = 0$$
$$TH_1(t+1) = -\Delta + \alpha TH_1(t) \quad \text{if } D_1 = 1 \text{,} \qquad (2)$$

where $\alpha$ is referred to as the forgetting (memory) parameter[12] and $\Delta$ is the constant increment (in this experiment, $\Delta$ = 1 and $\alpha$ = 0.99). The intuitive meaning of the update given by Eq. (2) is that the threshold value is revised so that the likelihood of choosing the *same* machine in the next cycle increases.

**(2) Second-most significant bit**

The threshold adjuster values related to the *second*-most significant bit are revised based on the following rules:

$$TH_{2,0}(t+1) = +\Delta + \alpha TH_{2,0}(t) \quad \text{if } D_1 = 0,\ D_2 = 0$$
$$TH_{2,0}(t+1) = -\Delta + \alpha TH_{2,0}(t) \quad \text{if } D_1 = 0,\ D_2 = 1 \qquad (3)$$

when the MSB has been determined to be 0 ($D_1 = 0$) and

$$TH_{2,1}(t+1) = +\Delta + \alpha TH_{2,1}(t) \quad \text{if } D_1 = 1,\ D_2 = 0$$
$$TH_{2,1}(t+1) = -\Delta + \alpha TH_{2,1}(t) \quad \text{if } D_1 = 1,\ D_2 = 1 \qquad (4)$$

when the MSB has been determined to be 1 ($D_1 = 1$).

**(3) General form**

As was done with the most and second-most significant bits in Eqs. (2)–(4), all of the threshold values are updated. In a general form, the threshold value for the *K*-th bit is given by

$$
\begin{aligned}
&TH_{K,S_1,S_2,\cdots,S_{K-1}}(t+1) = +\Delta + \alpha TH_{K,S_1,S_2,\cdots,S_{K-1}}(t) \quad \text{if } S_K = 0,\ D_1 = S_1,\ D_2 = S_2, \cdots,\ D_{K-1} = S_{K-1} \\
&TH_{K,S_1,S_2,\cdots,S_{K-1}}(t+1) = -\Delta + \alpha TH_{K,S_1,S_2,\cdots,S_{K-1}}(t) \quad \text{if } S_K = 1,\ D_1 = S_1,\ D_2 = S_2, \cdots,\ D_{K-1} = S_{K-1}
\end{aligned}
\qquad (5)
$$

when the decisions from the MSB to the ($K$-1)-th bit have been determined by $D_1 = S_1$, $D_2 = S_2$, ..., and $D_{K-1} = S_{K-1}$.

**II. When *losing* the slot machine play**

If the selected machine does *not* yield a reward (i.e. the player *loses* in the slot machine play), the threshold values are updated as follows.

**(1) MSB**

The threshold value of the MSB is updated according to

$$
\begin{aligned}
&TH_1(t+1) = -\Omega_1 + \alpha TH_1(t) \quad \text{if } D_1 = 0 \\
&TH_1(t+1) = +\Omega_1 + \alpha TH_1(t) \quad \text{if } D_1 = 1
\end{aligned},
\qquad (6)
$$

where the parameter $\Omega_1$ is determined based on the history of betting results.

Let the number of times that slot machines for which the MSB is 0 ($S_1 = 0$) and 1 ($S_1 = 1$) are *selected* be given by $C_{S_1=0}$ and $C_{S_1=1}$, respectively. Let the number of *wins* by selecting slot machines for which the MSB is 0 and 1 be given by $L_{S_1=0}$ and $L_{S_1=1}$, respectively. Then, the estimated reward probability, or winning probability, by choosing slot machines for which the MSB is $k$ is given by

$$
\hat{P}_{S_1=k} = \frac{L_{S_1=k}}{C_{S_1=k}}, \qquad (7)
$$

where $k$ is 0 or 1. $\Omega_1$ is then given by

$$\Omega_1 = \frac{\hat{P}_{S_1=0} + \hat{P}_{S_1=1}}{2 - (\hat{P}_{S_1=0} + \hat{P}_{S_1=1})}. \tag{8}$$

The initial $\Omega_1$ is assumed to be unity, and a constant value is assumed when the denominator of Eq. (8) is 0. $\Omega_1$ is the figure that designates the degrees of winning and losing. Indeed, the numerator of Eq. (8) indicates the degree of winning, whereas the denominator shows that of losing.

**(2) Second-most significant bit**

The threshold adjuster values related to the *second*-most significant bit are updated when the MSB of the decision is 0 ($D_1 = 0$) by using the following formula:

$$\begin{aligned} TH_{2,0}(t+1) &= -\Omega_{2,0} + \alpha TH_{2,0}(t) \quad \text{if } D_1 = 0,\ D_2 = 0 \\ TH_{2,0}(t+1) &= +\Omega_{2,0} + \alpha TH_{2,0}(t) \quad \text{if } D_1 = 0,\ D_2 = 1 \end{aligned}. \tag{9}$$

Let the number of times that slot machines for which the MSB is 0 ($S_1 = 0$) *and* the second-most significant bit is 0 ($S_2 = 0$) are *selected* be given by $C_{S_1=0,S_2=0}$. Let the number of times that slot machines for which the MSB is 0 ($S_1 = 0$) *and* the second-most significant bit is 1 ($S_2 = 1$) are selected be given by $C_{S_1=0,S_2=1}$. Let the numbers of *wins* by selecting slot machines for which the MSB is 0 ($S_1 = 0$) *and* the second-most significant bit is 0 ($S_2 = 0$) or 1 ($S_2 = 1$) be given by $L_{S_1=0,S_2=0}$ and $L_{S_1=0,S_2=1}$, respectively. Then, the estimated reward probability, or winning probability, by choosing slot machines *for which the MSB is 0 and* the *second-most significant bit* is $k$ is given by

$$\hat{P}_{S_1=0,S_2=k} = \frac{L_{S_1=0,S_2=k}}{C_{S_1=0,S_2=k}}. \tag{10}$$

$\Omega_{2,0}$ is then given by

$$\Omega_{2,0} = \frac{\hat{P}_{S_1=0,S_2=0} + \hat{P}_{S_1=0,S_2=1}}{2-(\hat{P}_{S_1=0,S_2=0} + \hat{P}_{S_1=0,S_2=1})}. \quad (11)$$

$\Omega_{2,0}$ concerns the ratios of winning and losing within the slot machine groups whose MSBs are given by 0.

### (3) General form

All of the threshold values are updated. In a general form,

$$\begin{aligned} TH_{K,S_1,S_2,\cdots,S_{K-1}}(t+1) &= -\Omega_{K,S_1,S_2,\cdots,S_{K-1}} + \alpha TH_{K,S_1,S_2,\cdots,S_{K-1}}(t) \quad \text{if } S_K = 0,\ D_1 = S_1,\ D_2 = S_2, \cdots,\ D_{K-1} = S_{K-1} \\ TH_{K,S_1,S_2,\cdots,S_{K-1}}(t+1) &= +\Omega_{K,S_1,S_2,\cdots,S_{K-1}} + \alpha TH_{K,S_1,S_2,\cdots,S_{K-1}}(t) \quad \text{if } S_K = 0,\ D_1 = S_1,\ D_2 = S_2, L,\ D_{K-1} = S_{K-1}. \end{aligned} \quad (12)$$

$\Omega_{K,S_1,S_2,\cdots,S_{K-1}}$ is the increment defined as follows.

Let the number of times that slot machines whose upper $K$-1 bits are specified by $S_1, S_2, \cdots, S_{K-1}$ and whose $K$-th bits are given by $k$ ($S_k = k$) are *selected* be denoted by $C_{S_1 S_2 \cdots S_{K-1}, S_K=k}$. Let the number of *wins* by selecting such machines be given by $L_{S_1 S_2 \cdots S_{K-1}, S_K=k}$. Then, the estimated reward probability, or winning probability, by choosing such machines is given by

$$\hat{P}_{S_1 S_2 \cdots S_{K-1}, S_K=k} = \frac{L_{S_1 S_2 \cdots S_{K-1}, S_K=k}}{L_{S_1 S_2 \cdots S_{K-1}, S_K=k}}. \quad (13)$$

$\Omega_{K,S_1,S_2,\cdots,S_{K-1}}$ is then given by

$$\Omega_{K,S_1,S_2,\cdots,S_{K-1}} = \frac{\hat{P}_{S_1 S_2 \cdots S_{K-1}, S_K=0} + \hat{P}_{S_1 S_2 \cdots S_{K-1}, S_K=1}}{2-(\hat{P}_{S_1 S_2 \cdots S_{K-1}, S_K=0} + \hat{P}_{S_1 S_2 \cdots S_{K-1}, S_K=1})}. \quad (14)$$

Initially, all of the threshold values are 0; hence, for example, the probability of determining the MSB of the slot machine to be 1 or 0 is 0.5 since $TH_1 = 0$. As time elapses, the threshold values are biased towards the slot machine with the higher reward probability based on the updates described by Eqs. (2)–(10). It should be noted that due to the irregular nature of the incoming chaotic signal, the possibility of choosing the *opposite* values of bits is *not* 0, and this feature is critical in exploration in reinforcement learning. For example, even when the value of $TH_1$ is sufficiently small (indicating that slot machines whose MSBs are 1 are highly likely to be better machines), the probability of the decision to choose machines whose MSBs are 0 is *not* 0.

The number of threshold levels was limited to a finite value in the experimental implementation. Furthermore, the threshold resolution affects the decision-making performance, as discussed below. In this study, we assumed that the actual threshold level takes the values $-Z, \ldots, -1, 0, 1, \ldots, Z$, where $Z$ is a natural number; thus, the number of the threshold levels is $2Z+1$ (Fig. 1c). More precisely, the actual threshold value is defined by

$$T(t) = a \times \lfloor TH(t) \rfloor, \tag{15}$$

where $\lfloor TH(t) \rfloor$ is the nearest integer to $TH(t)$ rounded to 0, and $a$ is a constant for scaling to limit the range of the resulting $T(t)$. The value of $T(t)$ ranges from $-aZ$ to $aZ$ by assigning the limits $T(t) = aZ$ when $\lfloor TH(t) \rfloor$ is greater than $Z$ and $T(t) = -aZ$ when $\lfloor TH(t) \rfloor$ is smaller than $-Z$. In the experimental results shown below, the chaotic signals $s(t)$ take integer values from $-127$ to $128$; hence, $a$ was given by $a = 128/Z$ in the present study.

## Data analysis

**(1) Chaotic and quasiperiodic time series:** Four kinds of chaotically oscillating signal trains (Chaos 1, 2, 3, and 4) and a quasiperiodically oscillated sequence were sampled at a rate of 100 GSample/s using 10,000,000 points, which took approximately 10 μs. Such 10 M points were stored 120 times for each signal train; hence, there were 120 kinds of 10-M-long sequences. The processing for decision making was performed offline using a personal computer (Hewlett-Packard, Z-800, Intel Xeon CPU, 3.33 GHZ, 48 GB RAM, Windows 7, MATLAB R2016b). Pseudorandom sequences were generated using the random number generator by Mersenne Twister implemented in MATLAB.

**(2) Coloured noise:** Coloured noise was calculated based on the Ornstein–Uhlenbeck process using white Gaussian noise and a low-pass filter in numerical simulations[25]. We assumed that the coloured noise was generated at a sampling rate of 100 GHz, and the cut-off frequency of the low-pass filter was set to 10 GHz. Forty sequences of 10,000,000 points were generated. The reduction in the number of sequences was due to the excessive computational cost.

**(3) Two-armed bandit problem shown in Fig. 2c:** Five hundred consecutive plays were repeated 10,000 times; hence, the total number of slot machine plays for a single chaotic sequence was 5,000,000. Such plays were performed for all 120 sets of sequences. (In the case of coloured noise, 40 sets of sequences were used.) The CDR described in the main text was derived as the average of these sets. In Fig. 2c, the first 250 plays are shown to highlight the initial adaptation.

**(4) Two-armed bandit problem shown in Fig. 3:** The number of consecutive plays was 100. All of the other conditions were the same as in **(3)**.

**(5) Multi-armed bandit demonstrations in Figs. 4-6:** The numbers of plays in the multi-armed bandit problems whose results are depicted in Figs. 4–6 are summarized in Table 1. The suppression of the number of repetitions is due to the limitations of our computing environment.

**Table 1** Setting of the slot machine playing

| Number of bandits | Number of consecutive plays | Number of repetitions |
|---|---|---|
| 2 | 500 | 10,000 |
| 4 | 1000 | 1000 |
| 8 | 5000 (shown only until 2000 in Fig. 4) | 100 |
| 16 | 5000 (shown only until 3000 in Fig. 4) | 100 |
| 32 | 5000 | 100 |
| 64 | 10,000 | 100 |

**(6) 32-armed bandit problem setting (Fig. 4a(v)):** The reward probabilities of Machines 0–31 were 0.7, 0.5, 0.9, 0.1, 0.7, 0.5, 0.7, 0.5, 0.7, 0.5, 0.7, 0.5, 0.7, 0.5, 0.7, 0.5, 0.7, 0.5, 0.7, 0.5, 0.7, 0.5, 0.7, 0.5, 0.7, 0.5, 0.7, 0.5, 0.7, 0.5, 0.7, and 0.5, where the best machine was Machine 2. The contradiction conditions were satisfied. For example, the MSB of the best machine (Machine 2) was 0, but the sum of the reward probabilities of the machines whose MSBs were 0 was smaller than that for the machines whose MSBs were 1 (9.4 versus 9.6); the same structure held from the second to fourth bits as well.

**(7) 64-armed bandit problem setting (Fig. 4a(vi)):** The reward probabilities of Machines 0–63 were 0.7, 0.5, 0.9, 0.1, 0.7, 0.5, 0.7, 0.5, 0.7, 0.5, 0.7, 0.5, 0.7, 0.5, 0.7, 0.5, 0.7, 0.5, 0.7, 0.5, 0.7, 0.5, 0.7, 0.5, 0.7, 0.5, 0.7, 0.5, 0.7, 0.5, 0.7, 0.5, 0.7, 0.5, 0.7, 0.5, 0.7, 0.5, 0.7, 0.5, 0.7, 0.5, 0.7, 0.5, 0.7, 0.5, 0.7, 0.5, 0.7, 0.5, 0.7, 0.5, 0.7, 0.5, 0.7, 0.5, 0.7, 0.5, 0.7, 0.5, 0.7, 0.5, 0.7, and 0.5, where the best machine was Machine 2. The contradiction conditions were satisfied similarly to in the 32-armed bandit problem described in **(6)**.

**(8) Autocorrelation of chaotic signals (Fig. 2b):** The autocorrelation was computed based on all 10,000,000 sampling points of each sequence. The autocorrelation in Fig. 2b is the average of 120 kinds of autocorrelations. (For the coloured noise, the number of sequences was 40.)

## Acknowledgements

This work was supported in part by CREST project (JPMJCR17N2) funded by the Japan Science and Technology Agency, the Core-to-Core Program A. Advanced Research Networks and the Grants-in-Aid for Scientific Research (A) (JP17H01277) and (B) (JP16H03878) funded by the Japan Society for the Promotion of Science.

## Author contributions

M.N. and A.U. directed the project. M.N. designed the system architecture and the principle. T.M. and A.U designed and implemented the laser chaos. T.M., A.U., and M.N. conducted the optical experiments and data processing. M.N., T.M., A.U., H.H., H.S., and K.O. analysed the data. M.N. and A.U. wrote the paper.

## Competing interests

The authors declare no competing interests.

**Correspondence and requests for materials** should be addressed to M.N.

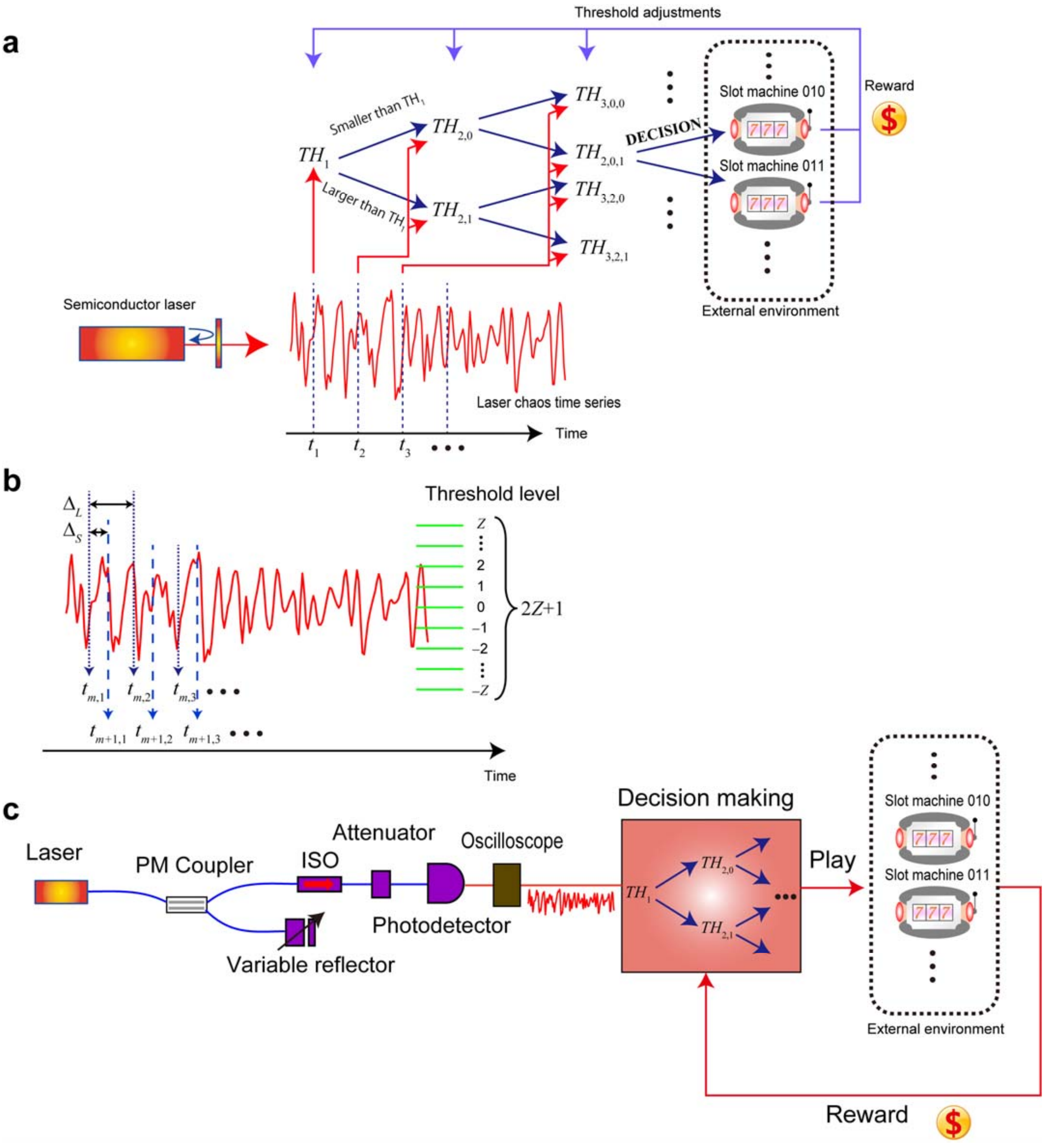

**Fig. 1 | Architecture for scalable reinforcement learning using laser chaos. a** Solving the multi-armed bandit problem with $N = 2^M$ arms using a pipelined arrangement of comparisons between thresholds and a series of chaotic signal sequences. **b** Chaotic time series with the definitions of the

inter-decision sampling interval ($\Delta_S$) and inter-bit sampling interval ($\Delta_L$) to arrive at a single decision. The $2Z+1$ threshold levels are also depicted, where $Z$ is a natural number. **c** Schematic diagram of the decision-making system architecture based on laser chaos and pipelined threshold processing.

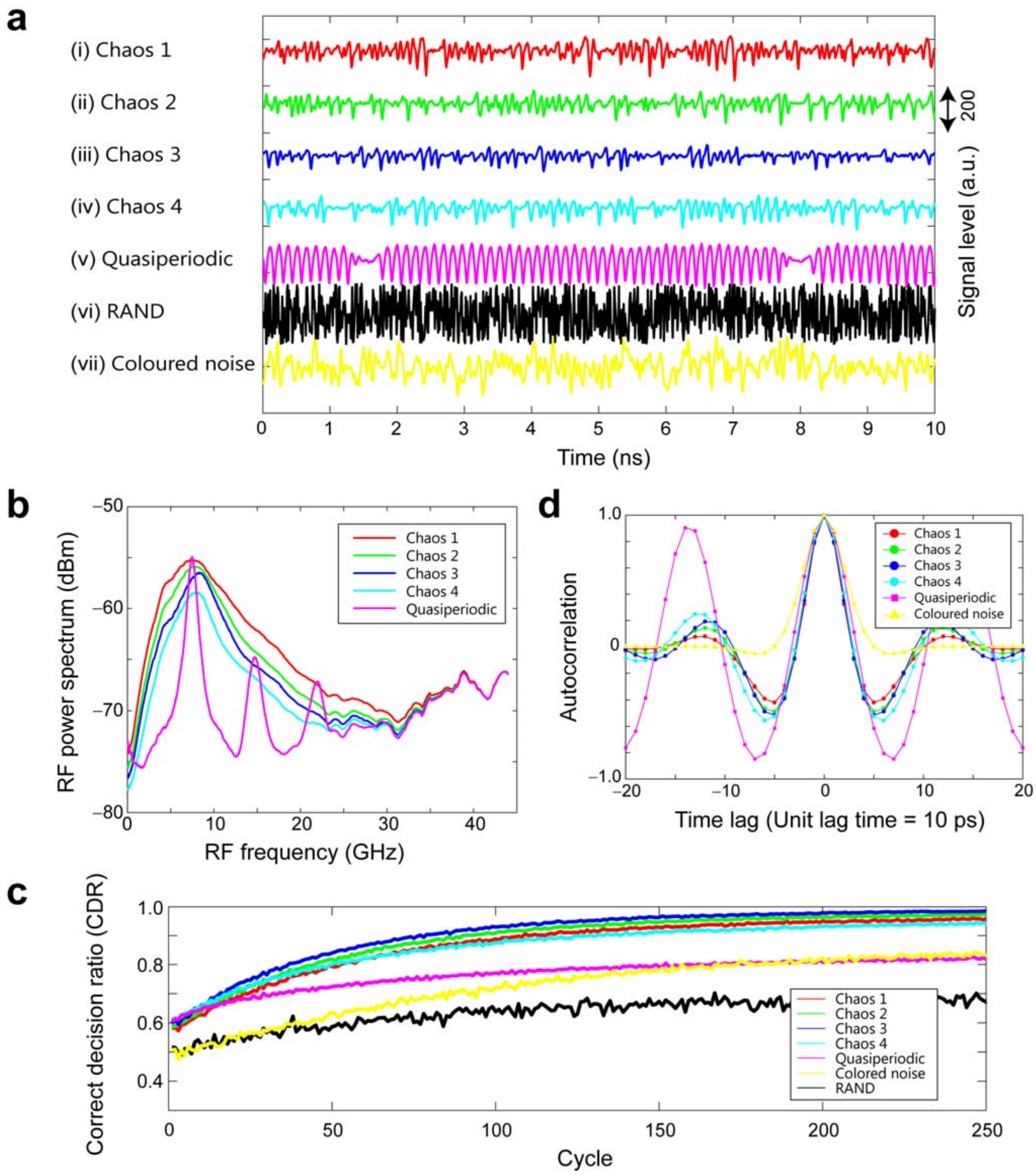

**Fig. 2 | Chaotic time series, inherent time-correlated structures, RF spectra, and decision-making performance for the two-armed bandit problem. a** Snapshots of the time series used for solving the *N*-armed bandit problem. Four kinds of chaotic signals (Chaos 1–4) as well

as quasiperiodic sequences, pseudorandom numbers (RAND), and coloured noise are used. **b** Radio-frequency (RF) power spectra in Chaos 1–4 and quasiperiodic signal cases. **c** Evolution of the correct decision ratio (CDR) indicating the likelihood of choosing the highest-reward-probability slot machine. **d** Autocorrelation inherent in Chaos 1–4, quasiperiodic, and coloured noise cases.

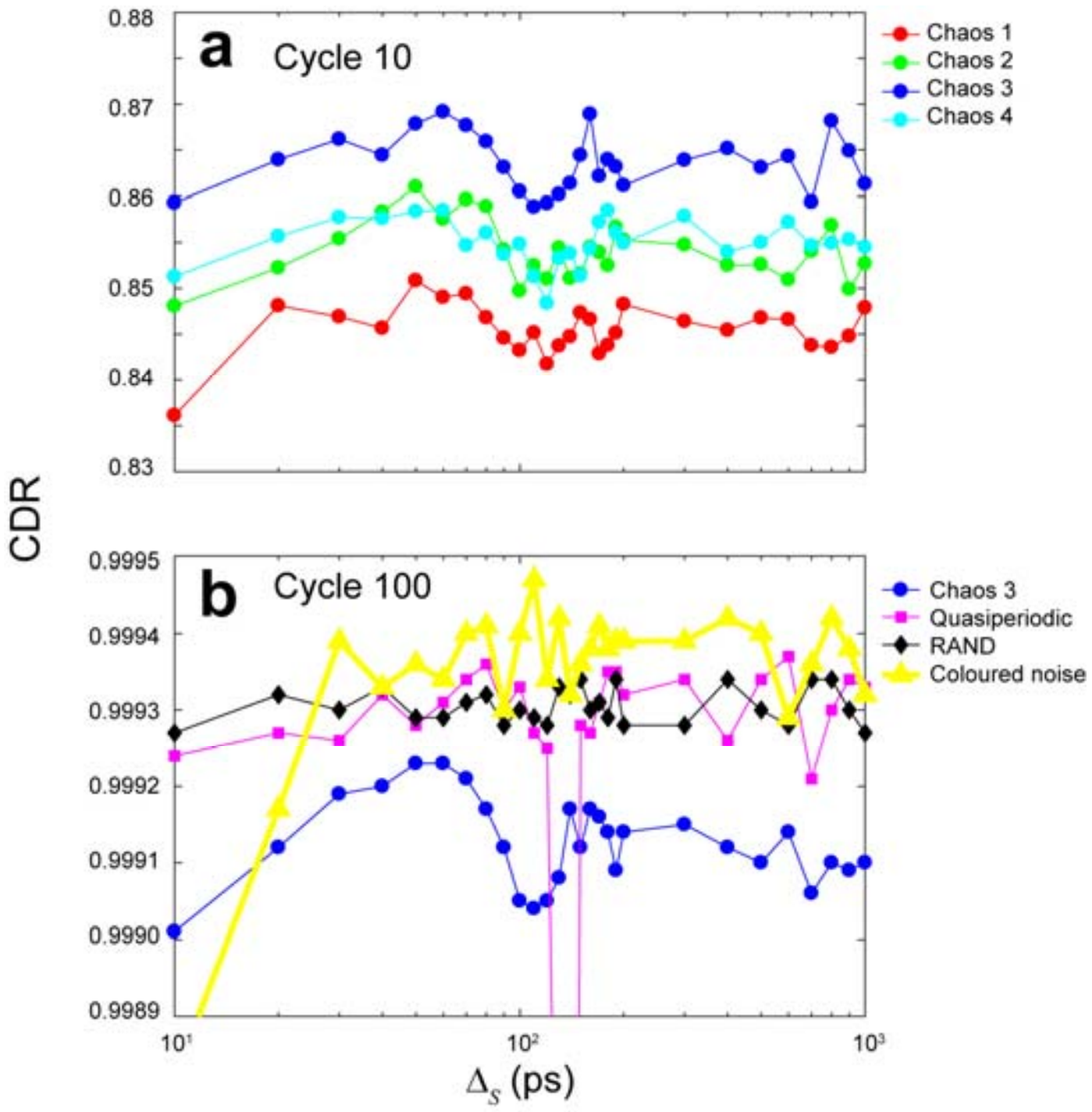

**Fig. 3 | Decision-making performance as a function of inter-decision sampling interval.**

CDR comparison at cycles **a** 10 and **b** 100.

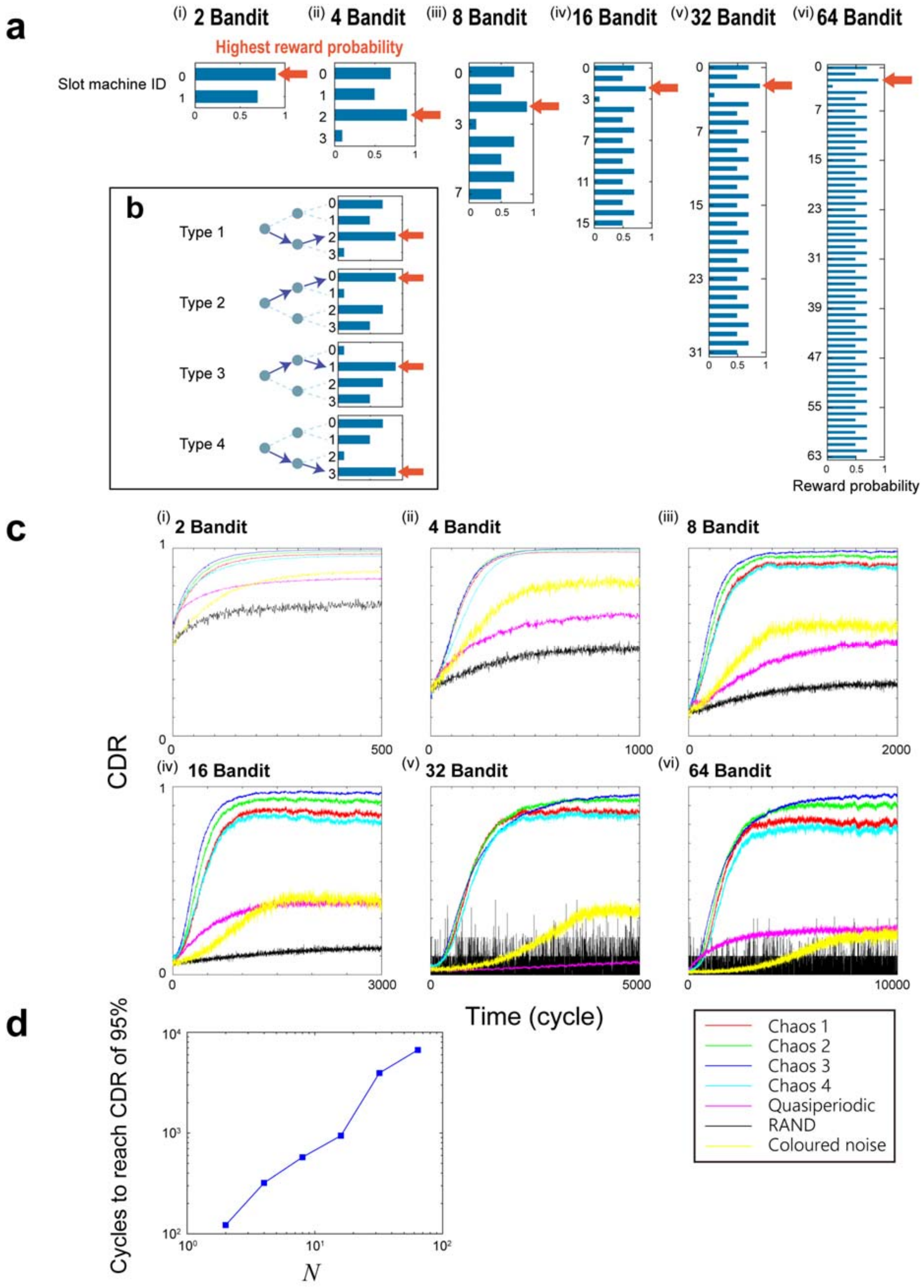

**Fig. 4 | Scalable decision making of *N*-armed bandit problems. a** Definitions of the two-, four-, eight-, 16-, 32, and 64-armed problems. The conditions are equally arranged to ensure fair

comparisons. **b** Four kinds of reward probability arrangements in the four-armed bandit problem. **c** CDR evolution. **d** Number of cycles necessary to reach a CDR of 0.95 as a function of $N$, which varies approximately as $N^{1.16}$.

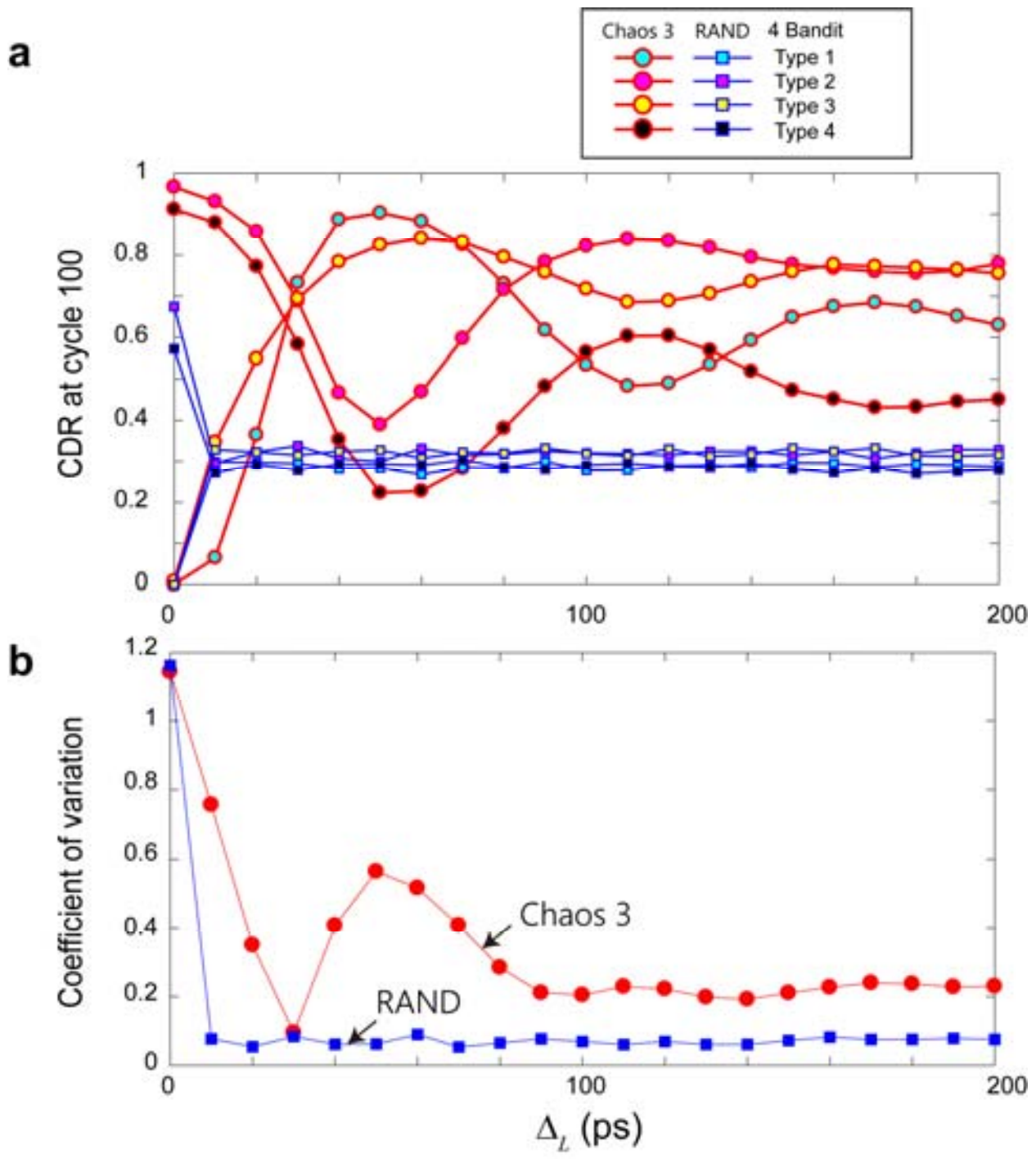

**Fig. 5 | Inter-bit sampling interval dependency. a** Effects of the temporal structure inherent in laser chaos on the decision-making performance, which can be negative or positive depending on the given problem. **b** Evaluation of a moderate inter-bit sampling interval choice based on the coefficient of variation (CV) of **a**. $\Delta_L$ = 30 ps or 100 ps yields a lower CV for the Chaos 3 case. Meanwhile, the unstructured time-domain properties of pseudorandom numbers are clearly shown by the flat dependency on $\Delta_L$ in the RAND case.

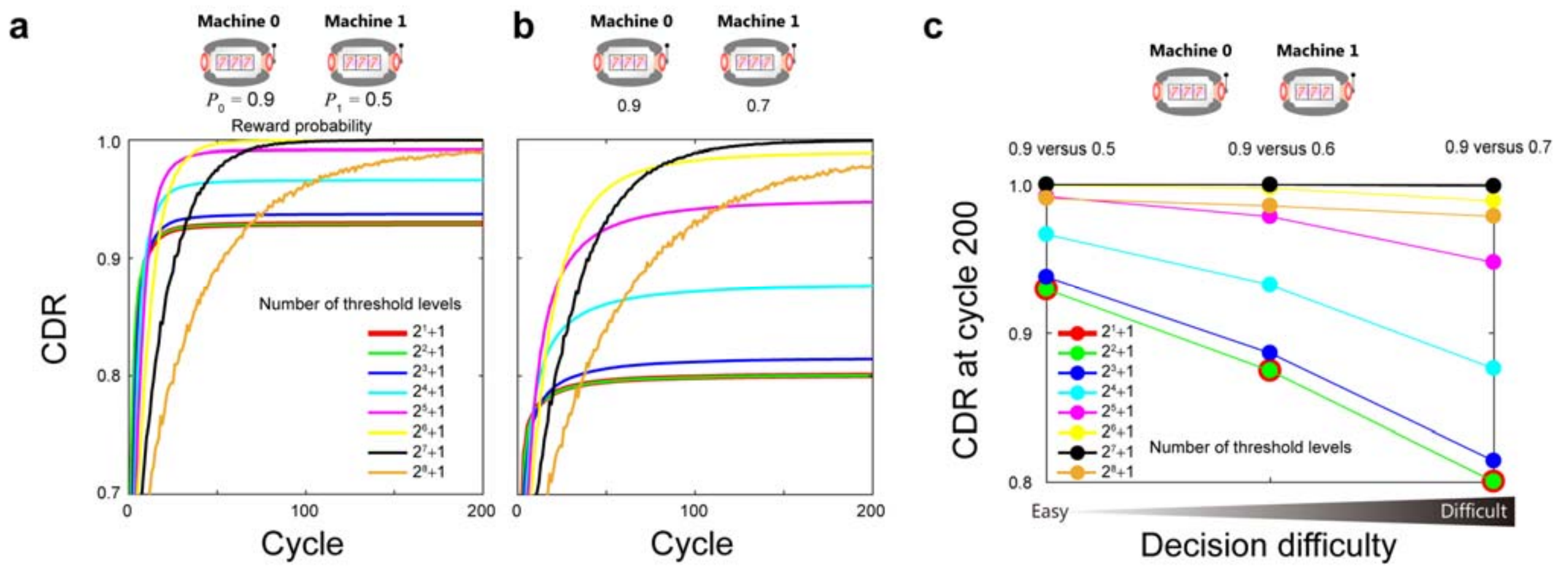

**Fig. 6 | Difficulty of the given decision-making problem and threshold precision control.**

While retaining the higher reward probability in the two-armed bandit problem ($P_0$ = 0.9), the lower reward probability $P_1$ was set to **a** 0.5 and **b** 0.7 to examine the decision difficulty. The CDR increases more rapidly in the easier decision-making problem (**a**) than in the harder one (**b**). In addition, a decrease in the number of threshold levels prevents the system from reaching the correct decision, especially for harder problems, due to insufficient exploration. **c** CDR at cycle 200 as a function of decision difficulty.

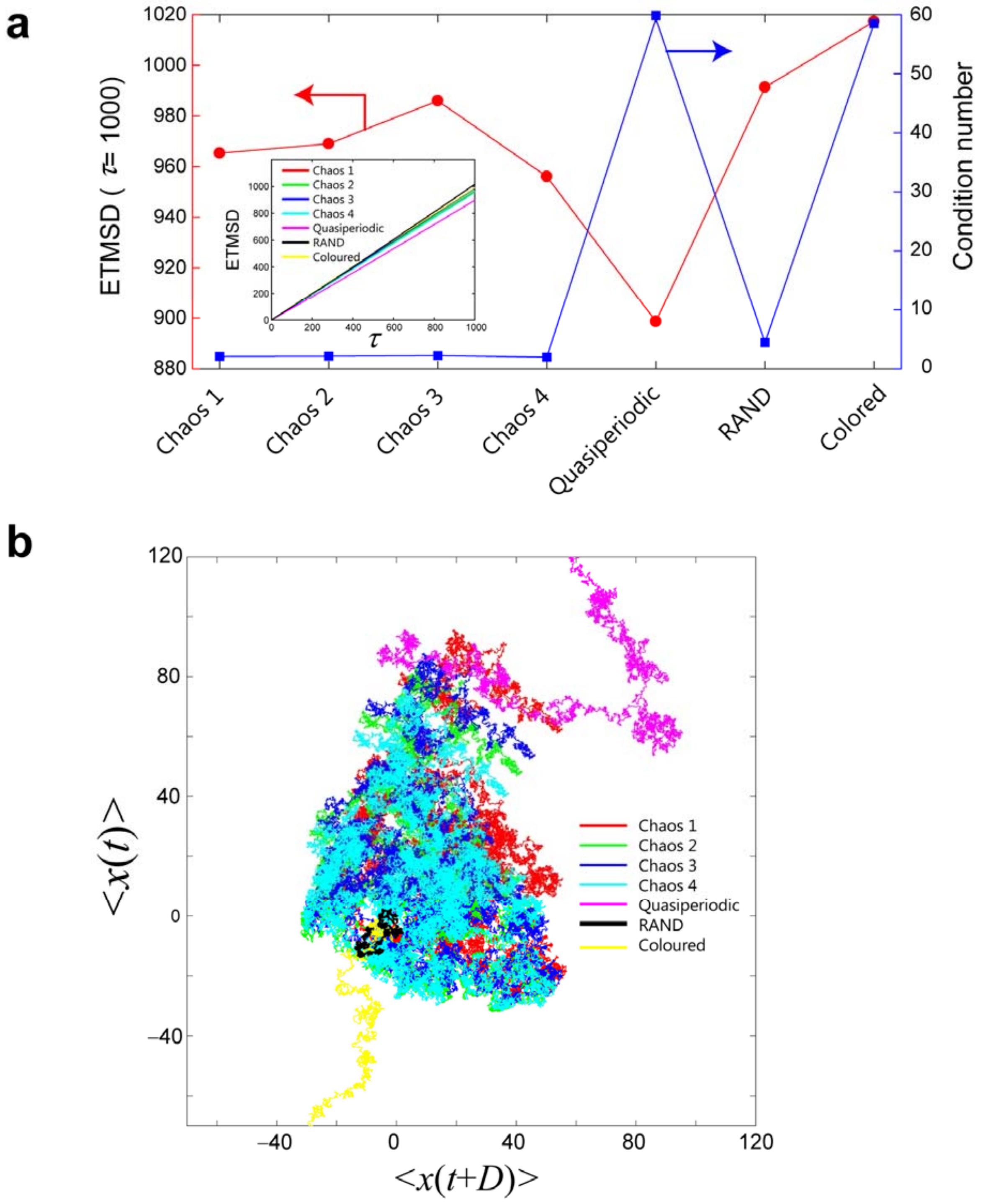

**Fig. 7 | Diffusivity analysis of the time series to investigate the underlying structure. a** Comparison of the ensemble averages of the time-averaged mean square displacements (ETMSDs), where the values for Chaos 1–4 agree with the superiority of the resulting decision-making performance. **b** Trajectories configured in a two-dimensional plane by $\langle x(t)\rangle$ and $\langle x(t+D)\rangle$ with $D$

= 10,000 to examine the coverage by the time series. The trajectories spanned by the chaotic time series are uniformly distributed and are also quantitatively analysed using the smaller condition numbers shown in **a**.